\documentclass[10pt]{article}

\usepackage{authblk}
\usepackage{amssymb}
\usepackage{amsmath}
\usepackage{algorithm}
\usepackage{algpseudocode}
\usepackage{subcaption}
\usepackage{graphicx}
\usepackage{multicol}
\usepackage{layouts}
\usepackage{pgf}
\usepackage{textcomp}
\usepackage{cite}
\def\BibTeX{{\rm B\kern-.05em{\sc i\kern-.025em b}\kern-.08em
    T\kern-.1667em\lower.7ex\hbox{E}\kern-.125emX}}
\usepackage{fancyhdr}
\usepackage[verbose=true,letterpaper]{geometry}
\AtBeginDocument{
  \newgeometry{
    textheight=9in,
    textwidth=6.5in,
    top=1in,
    headheight=14pt,
    headsep=25pt,
    footskip=30pt
  }
}

\title{SHIELD: Secure Haplotype Imputation Employing Local Differential Privacy}
\author[1]{Marc Harary\thanks{marc.harary@yale.edu}}
\affil[1]{Yale University}

\begin{document}

\maketitle


\begin{abstract}
We introduce Secure Haplotype Imputation Employing Local Differential privacy (SHIELD), a program for accurately estimating the genotype of target samples at markers that are not directly assayed by array-based genotyping platforms while preserving the privacy of donors to public reference panels. At the core of SHIELD is the Li-Stephens model of genetic recombination, according to which genomic information is comprised of mosaics of ancestral haplotype fragments that coalesce via a Markov random field. We use the standard forward-backward algorithm for inferring the ancestral haplotypes of target genomes---and hence the most likely genotype at unobserved sites---using a reference panel of template haplotypes whose privacy is guaranteed by the randomized response technique from differential privacy.
\end{abstract}

\section{Introduction}

In the context of biomedical analyses of large patient cohorts, whole-genome sequencing still remains prohibitively expensive for existing high-throughput technology. On the other hand, array-based genotyping platforms provide a more efficient method of collecting data for large-scale studies of human disease, albeit at the expense of the statistical power of genome-wide association (GWA) studies that intend to fine-map causal variants or facilitate meta-analyses \cite{international2007second, marchini2007new, servin2007imputation, taliun2021sequencing, li2009genotype}.

One solution is genotype imputation, a preliminary stage in many GWA studies that consists of inferring the genotype for a given target genome at loci that have not been directly assayed, essentially expanding the dimensionality of the original dataset \cite{marchini2010genotype, li2009genotype, li2010mach, das2016next, ayres2012beagle, purcell2007plink, marchini2007new}. Employing a reference panel of donated haplotypes sequenced via higher-quality technology and at a far denser set of variants, imputation algorithms like MaCH \cite{li2010mach}, Minimac \cite{das2016next}, BEAGLE \cite{ayres2012beagle}, PLINK \cite{purcell2007plink}, fastPHASE \cite{scheet2006fast}, and IMPUTE \cite{marchini2007new} have been demonstrated to reliably augment both the coverage and statistical power of GWA analyses and hence become an essential component of many clinical studies \cite{marchini2010genotype}.

Further to this end, public databases like the UK biobank (UKB) \cite{sudlow2015uk}, All of Us research program \cite{all2019all}, Haplotype Reference Consortium \cite{haplotype2016reference}, and 1,000 Genomes Project (1KG) \cite{siva20081000} have been made available to facilitate genomic research in part by offering standardized and readily accessible reference panels \cite{das2018genotype}. In cases where running imputation algorithms using large reference panels is impractical on local hardware or the direct access to the biobank data is prohibited, public web services like the Michigan Impute Server \cite{loh2016reference} are often established to answer queries to clients submitting target haplotypes for imputation.

Unfortunately, as part of a growing literature on privacy concerns in genomic research, it has also been documented that coordinated attacks on the part of cryptographic adversaries are capable of compromising the privacy of research subjects that donate to public reference panels \cite{dokmai2021privacy, bonomi2020privacy, sankararaman2009genomic, homer2008resolving}. For example, attackers have been able to exploit ancestral data \cite{erlich2018identity} or other personally identifying information \cite{gymrek2013identifying} to reconstruct reference genomes. An urgent challenge is therefore to develop a suite of imputation algorithms that can simultaneously facilitate high-utility, statistically reliable GWA studies while protecting the privacy of contributors to reference haplotype panels \cite{cho2018secure, dokmai2021privacy, naveed2015privacy}.

One solution is the technique of differential privacy, which has rapidly become the ``gold-standard'' for statistical queries by being able to provide both robust privacy guarantees for participants in studies and meaningful results for researchers in commercial and scientific settings \cite{dwork2006differential, dwork2008differential, dankar2013practicing}. At the crux of the technique is a rigorous mathematical formalization of privacy that quantifies the extent to which adding pseudorandom noise to the results of computations can protect the anonymity of members of a database \cite{dwork2014algorithmic}. 

The following work introduces Secure Haplotype Imputation Employing Local Differential privacy (SHIELD), a program that employs the Li-Stephens model of genetic recombination \cite{li2003modeling, li2009genotype} to impute missing haplotype variants in target genomes while incorporating differential privacy techniques to protect reference panel donors. Specifically, SHIELD proceeds in two stages: (i) initial input perturbation to guarantee local differential privacy \cite{yang2020local} via randomized response \cite{wang2016using, warner1965randomized} and (ii) fitting a hidden Markov model \cite{rabiner1986introduction} to each subsequent client query via the forward-backward algorithm \cite{baum1972inequality}. In an experiment that closely simulates a real-world use case for haplotype imputation, we show that SHIELD is able to obtain state-of-the-art imputation accuracy while providing mathematically formalized privacy guarantees.

\section{Results}

\subsection{Overview}

The setting for which SHIELD is intended consists of a client user uploading target genomes to a public imputation server \cite{dokmai2021privacy}. In the standard imputation workflow, contributors to a biobank upload their sequenced genomic data to a central, publicly available server, where the data are then collated to create a haplotype reference panel to pass as an argument to an imputation algorithm \cite{sudlow2015uk, siva20081000, haplotype2016reference}. Subsequently, client researchers may then upload target genomes as part of a clinical study to the server, where the targets are imputed using the private haplotype reference panel and, most often, an algorithm based in hidden Markov models \cite{rabiner1986introduction, li2010mach, das2016next, ayres2012beagle, purcell2007plink, marchini2007new} and the forward-backward algorithm \cite{baum1972inequality}. At no point in the workflow is the haplotype reference panel directly visible to client researchers submitting jobs to the server. However, while the privacy of the contributors to the reference panel may appear guaranteed, it has been demonstrated that adversarial attacks employing carefully coordinated queries to the server can divulge the sequences of reference haplotypes \cite{dokmai2021privacy}.

To this end, SHIELD modifies the imputation workflow by leveraging local \cite{yang2020local} differential privacy \cite{dwork2006differential, dwork2008differential, dankar2013practicing, cormode2018privacy, wang2016using}. Haplotype data can be represented as a bitstring in which a 1 at the $i$th position in the sequence indicates that the haplotype possesses the minor allele at the $i$th site and a 0 the major allele \cite{das2016next}. Prior to submission to the central imputation server, pseudorandom noise is added to the two bitstrings denoting each individual's pair of haplotypes via randomized response, a technique from differential privacy that simply consists of flipping a random subset of the bits from 0 to 1 and vice versa \cite{wang2016using, warner1965randomized}. The likelihood that a given bit in the haplotype bitstring is flipped varies as a function of a parameter $\varepsilon$---called the privacy budget \cite{dwork2014algorithmic}---such that lower values of $\varepsilon$ entail a higher probability that any bit is flipped and therefore a higher degree of privacy. The tradeoff, however, is that lower privacy budgets incur a greater expense to imputation accuracy, rendering it a hyperparameter that the database curator must carefully adjust to strike an acceptable balance between donor privacy and client utility. Once all perturbed haplotypes are collected at the central server, imputation is subsequently performed using the modified haplotypes as a reference panel.

Privacy is guaranteed by the fact that no contributor's data will, on average, be unmodified when input to the imputation algorithm invoked by client researchers. In this way, no adversary could be certain that the results that they obtain from an attack accurately reflect the true reference panel. These privacy guarantees are also local; even if an adversary were to access the reference panel directly rather than through coordinated queries, the data obtained would again not perfectly reflect any individual's true genome \cite{cormode2018privacy}.

\begin{figure}[t!]
	\centering
		\includegraphics{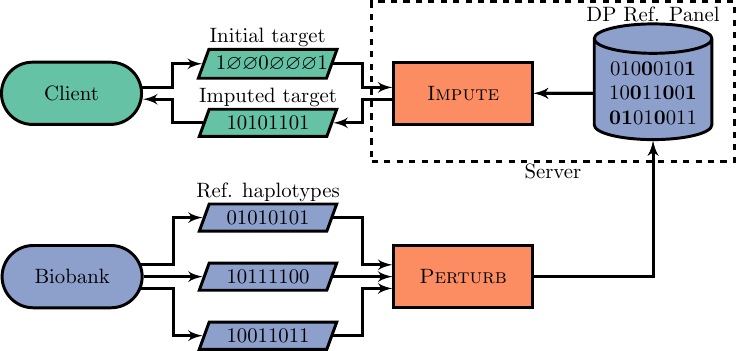}
		\caption{Overview of the SHIELD pipeline, with the key algorithms in orange. Noise is added once to the reference data (purple) via \textproc{Perturb}, then collated and stored on the server to guarantee local DP (modified bits in bold). The client (green) then calls \textsc{Impute} on the server with the target haplotype (missing sites denoted $\varnothing$) and the reference panel as arguments.}
\end{figure}

\subsection{State-of-the-art imputation accuracy}

\begin{figure}[t]
	\centering
		\includegraphics[width=1.\linewidth]{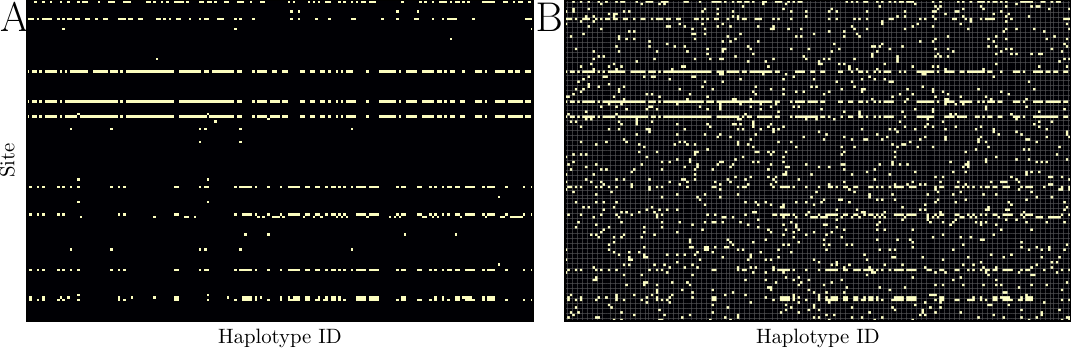}
		\caption{A. The reference haplotype matrix corresponding to the first 128 SNPs on chromosome 20 and 200 haplotypes in 1KG. Empty squares represent the presence of the major allele, yellow of the minor. B. The same haplotype matrix perturbed by SHIELD.}
\end{figure}



To evaluate SHIELD's performance on a realistic simulation of an imputation query, we performed an ablation study on the 1KG Phase 3 \cite{siva20081000} dataset. We withheld 100 genomes (equivalent to 200 haplotypes) from the reference panel to impute via the remaining 2,404 samples. The first 10,000 single-nucleotide polymorphisms (SNPs) were extracted from 1KG; the remaining were discarded to render run times more tractable. To simulate an array-based assay of the 200 target haplotypes, we ablated all sites except those included in the Illumina Human1M-Duo v3.0 DNA Analysis BeadChip manifest, the intersection of which with the first 10,000 sites in the 1KG data consisted of a total of 253 sites for an \textit{a priori} coverage of 2.53\%. 

To quantify accuracy, we summed the imputed dosages for each pair of haplotypes to compute a final genotype dosage for each sample, then computed the coefficient of determination ($R^2$) between the genotype dosages and the ground-truth exome data. Because sites vary massively by minor allele frequency (MAF), the loci were divided into three bins corresponding to MAFs of $(0\%, 0.5\%)$, $\left[ 0.5\%, 5\% \right)$, and $\left[ 5\%, 50\% \right]$. Respectively, these bins contained 5,943, 2,157, and 1,900 variants in the reference set. Accuracy was assessed, by bin, both to compare the performance of SHIELD to that of Minimac3 \cite{das2016next} and to characterize the effect of the privacy budget on our method's accuracy.

\begin{figure*}[t!]
	\centering
		\includegraphics[width=1.\textwidth]{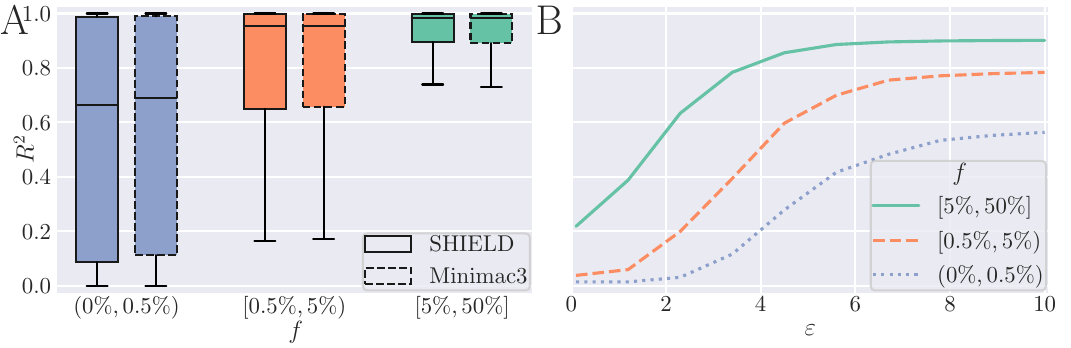}
		\caption{A. Comparison between the accuracy by MAF of imputed dosages for targets withheld from 1KG for both SHIELD (non-differentially private) and Minimac3. B. SHIELD's accuracy by MAF versus privacy budget.}
\end{figure*}

Our analyses show nearly identical performance between SHIELD and Minimac3 when no input perturbation is applied, with the former obtaining scores of 0.571, 0.784, and 0.902, respectively, on the three bins enumerated above and the latter scores of 0.584, 0.787, and 0.901 (Figure 2). SHIELD's accuracy was reevaluated at various values of our privacy budget along the interval $\left[0.01, 10 \right]$, reflecting the typical range of values that $\varepsilon$ is assigned in many differentially private algorithms \cite{dwork2006differential}. Expectedly, accuracy exhibits a negative association with $\varepsilon$. At an upper bound of $\varepsilon=10$, SHIELD performs nearly identically to Minimac3 (0.564, 0.784, 0.901; Figure 3), while performance degrades significantly at $\varepsilon=0.01$ (0.014, 0.038, 0.218; Figure 3).

%

\subsection{Impact on Markov parameters}

As noted above, the parameters for the Markov random field \cite{rabiner1986introduction} modeling genomic recombination \cite{li2003modeling}, namely the mutation and recombination rates, were computed on the unperturbed data by Minimac3 \cite{das2016next}. The rationale was that the noise added to the reference panel mimicked the behavior of extremely rapid genomic recombination, causing Minimac3's expectation-maximization procedure to dramatically overestimate the recombination rates (5.93 $\times 10^{-3}$ vs. 4.84$\times 10^{-4}$) and, conversely, to underestimate the mutation rates (Figure 3B). These atypical rates exerted a decidedly negative impact on imputation accuracy, with performance decreasing by 35.5\%, 16.1\%, and 5.46\% for each of the three bins, respectively, when the rates were computed on the reference panel perturbed at $\varepsilon=5.0$. In sum, it is clearly superior to estimate population parameters \textit{a priori}, although, notably, doing so on the reference panel itself is not differentially private and may leak information.

\subsection{Impact on compression rates}

An additional feature of haplotype imputation introduced by Minimac3 was the M3VCF format for genomic data, which both substantially decreases total file size over the traditional VCF format and enables the state-space reduction technique that further improves imputation runtime \cite{das2016next}. The key insight enabling the format is the observation that, due to identity-by-descent \cite{li2009genotype}, most haplotypes share identical $k$-mers of genomic material at intervals of contiguous loci despite being unique overall. In other words, given an arbitrary interval along the genome, the number of unique $k$-mers collectively exhibited by the reference panel is almost always smaller than the total number of reference haplotypes \textit{per se}. Therefore, it is possible to implement a compression scheme in which the genome is partitioned into intervals and only the unique $k$-mer strings are retained, substantially compressing the original reference panel \cite{das2016next}.

An unfortunate consequence of local differential privacy via randomized response is that, on average, random noise will destroy the exact equality between haplotypes substrings. From the perspective of a compression algorithm attempting to identify the set of unique $k$-mers along a given interval, an apparently larger number of unique fragments will exist, rendering M3VCF-style compression will less efficient. As an illustration, we partitioned the genomic data into mutually exclusive, exhaustive blocks of uniform size ranging from 2 to 500. We then computed the data compression ratio when M3VCF-style state-space reduction was applied at each block size by dividing the total $5.008 \times 10^8$ bits in the uncompressed panel by the number of bits following compression and plotted the ratio against block size (Figure 3C). Input perturbation resulted in compression rates up to an order of magnitude smaller.


\begin{figure*}[t]
	\centering
		\includegraphics[width=1.\textwidth]{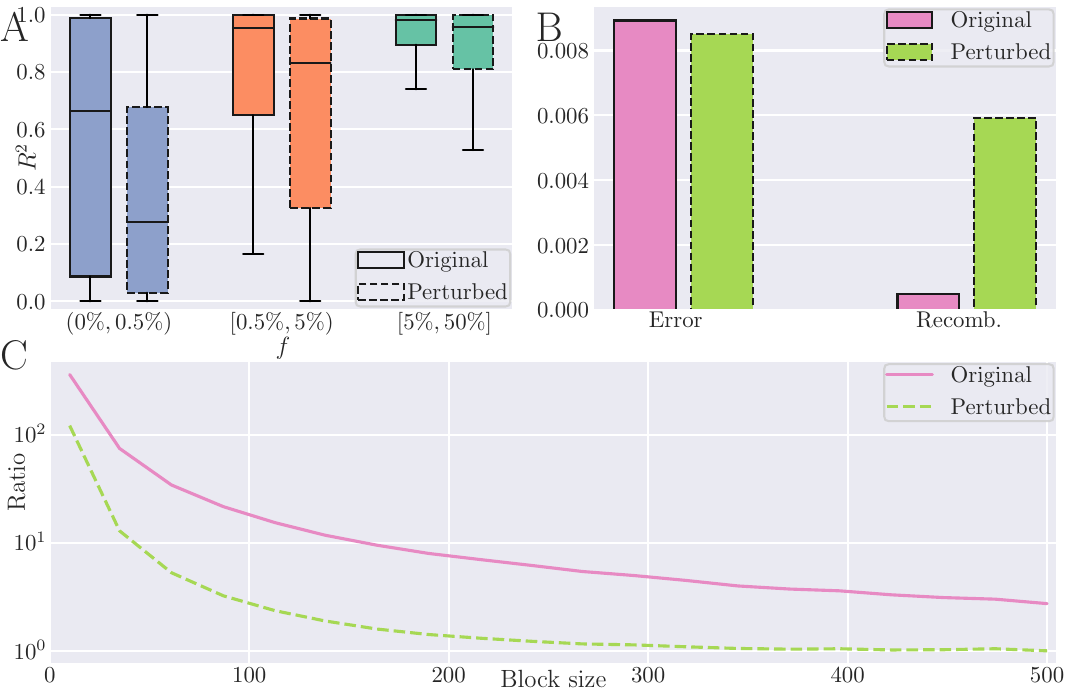}
		\caption{A. SHIELD's imputed dosage accuracy $\left( R^2 \right)$ by MAF using parameters derived from both the original and perturbed reference panels. B. The mean recombination and error rates using the original and perturbed parameters. C. M3VCF-like compression ratio versus haplotype block size on the original and perturbed parameters.}
\end{figure*}

\section{Discussion}

In this work, we develop Secure Haplotype Imputation Employing Local Differential privacy (SHIELD), a program for performing genomic imputation with strong privacy guarantees for reference haplotypes via the randomized response technique \cite{warner1965randomized}. Analysis shows that SHIELD is able to obtain state-of-the-art accuracy in realistic experimental settings at typical privacy budgets. 

We note that the strong performance of SHIELD parallels the effectiveness of RAPPOR \cite{erlingsson2014rappor}, a differentially private algorithm for mining strings in commercial contexts that is also based on randomized response. Unlike SHIELD, however, RAPPOR is not intended for data that is inherently binary; rather, arbitrary alphanumeric strings are hashed onto Bloom filters \cite{bloom1970space} that are subsequently perturbed. The fact that haplotype data intrinsically consist of bitstrings makes randomized response particularly convenient in a genomic context.

But despite the strong performance exhibited in the experiments above, it should be acknowledged that the privacy guarantees made by our program are limited to individual variants. In other words, for a given privacy budget $\varepsilon$ \cite{dwork2006differential, dwork2008differential, dankar2013practicing}, SHIELD can provably ensure protection for each sample's genotype at any one site, but not across the entire genome \textit{per se}. Certain adversarial attacks are therefore still feasible with SHIELD even though accurate reconstruction of reference haplotypes is not \cite{bonomi2020privacy, sankararaman2009genomic, homer2008resolving, erlich2018identity, gymrek2013identifying}. Whole-genome privacy would instead require the division $\varepsilon$ across each site (see \cite{dwork2008differential} for a discussion on composition in differential privacy), which is prohibitively difficult for datasets containing tens of thousands of variants. On the other hand, such divisions may be possible if a fairly limited segment of the genome is to be imputed. Future research into genomic privacy may investigate these scenarios or alternative differentially private mechanisms.

A second limitation of our program is its dependence on accurate \textit{a priori} estimates of population parameters \cite{li2003modeling, li2009genotype, das2016next}, which are non-trivial to compute while still enforcing local differential privacy. Subsequent work may inquire into the feasibility of computing population parameters \textit{a posteriori} by performing some manner of statistical correction. 

Nevertheless, the capacity for basic differentially private mechanisms to easily provide meaningful results is highly promising for the prospect of privacy in practical genomic research.

\section{Methods}

The SHIELD algorithm consists of two subroutines, \textproc{Perturb} and \textproc{Impute}, that are described below. The former is called once on a reference panel $\mathbf X$ to produce a locally \cite{yang2020local} differentially private \cite{dwork2006differential, dwork2008differential, dankar2013practicing, dwork2014algorithmic} reference panel $\mathbf{\tilde X}$ that is stored on the imputation server, whereas the latter is then called by the client for each subsequent query haplotype $\mathbf z$ using $\mathbf{\tilde X}$ as the reference panel.

\subsection{Differential Privacy and Randomized Response}
We derive the privacy guarantees of SHIELD from the notion of differential privacy \cite{dwork2006differential, dwork2008differential, dankar2013practicing, dwork2014algorithmic}. Preliminarily, we develop the notion of \textit{neighboring datasets}. Given a universe of datasets $\mathcal X$, we say that two datasets $x, y \in \mathcal X$ are neighbors if and only if they differ by at most one individual sample. We will also call a randomized algorithm $\mathcal{M}: \mathcal X \to \mathcal F$, where $\mathcal F$ is an arbitrary probability space, a \textit{mechanism}. We then say that a mechanism $\mathcal M: \mathcal X \to \mathcal F$ satisfies $\left( \epsilon, \delta \right)$-differential privacy if and only if for all $\mathcal S \subseteq \mathcal F$ and for all $x, y \in \mathcal X$ such that $x$ are $y$ are neighboring, we have
\begin{equation}
	P \left( \mathcal M\left(x\right) \in \mathcal S \right) \leq \exp \left(\varepsilon \right) P \left( \mathcal M\left(y\right) \in \mathcal S \right) + \delta.
\end{equation}

Among the most common techniques in differential privacy, randomized response \cite{wang2016using, warner1965randomized} satisfies $\epsilon$-differential privacy for binary attributes. The randomized response scheme on a binary attribute $X$ is a mechanism $\mathcal M_{rr}: \left\{0, 1 \right\} \to \left\{0, 1 \right\}$ is characterized by a $2 \times 2$ distortion matrix 
\begin{equation}
	\mathbf P = \begin{pmatrix} p_{00} & p_{01} \\ p_{10} & p_{11} \end{pmatrix},
\end{equation}
where $p_{uv} = P \left( \mathcal M_{rr}(x_i) = u | x_i = v \right) \quad \left( u, v \right) \in \left\{ 0, 1 \right\}$. It can be shown \cite{wang2016using} that the highest-utility value for $\mathbf P$ is 
\begin{equation}
	\mathbf P = \begin{pmatrix} \frac{e^\varepsilon}{1 + e^\varepsilon} & \frac{1}{1 + e^\varepsilon} \\ \frac{1}{1 + e^\varepsilon} & \frac{e^\varepsilon}{1 + e^\varepsilon} \end{pmatrix}.
\end{equation}

Fixing the number of samples in our reference panel $n$ and the number of sites $m$, we denote the universe of possible reference panels $\mathcal X = \left\{ 0, 1 \right\}^{m \times n}$. Because haplotypes are vector-valued, applying the notion of neighboring datasets is non-trivial. For our purposes, we will say that two reference panels $\mathbf X, \mathbf X^\prime \in \mathcal X$ are neighboring if and only if their Hamming distance is less than or equal to $1$. In other words, we consider $\mathbf X$ and $\mathbf{X^\prime}$ neighbors if and only if $X_{i,j} \neq X^\prime_{i,j}$ for a single marker $i$ and a single individual $j$ as opposed to a whole-genome interpretation of neighboring datasets in which $\mathbf X$ and $\mathbf X^\prime$ may differ by an entire row.

It then follows that by applying the randomized response mechanism $\mathcal M_{rr}$ to each entry in a reference panel matrix $\mathbf X$, we may store a perturbed copy $\mathbf{\tilde X}$ of the original reference panel that satisfies entry-wise $\varepsilon$-differential privacy. The perturbation step of SHIELD then consists of the procedure \textproc{Perturb}. We note that we use the symbol $\overset{r}{\gets}$ to denote a pseudorandom sample and $\text{Bern}\left( \vartheta \right)$ to denote a Bernoulli distribution with parameter $\vartheta$.
\begin{algorithm}[h!]
\caption{Applies randomized response mechanism to reference panel.}
\begin{algorithmic}[1]
\Procedure{Perturb}{$\mathbf X, \varepsilon$}
\State $\mathbf{\tilde X} \gets$ empty matrix
\For{$i = 1, 2, \ldots, n $}
	\For{$j = 1, 2, \ldots, m $}
		\State $c \overset{r}{\gets} \text{Bern} \left( \frac{e^\varepsilon}{1 + e^\varepsilon} \right)$
		\If {$c = 1$}
			\State $\tilde X_{i, j} \gets X_{i,j}$
		\Else
			\State $\tilde X_{i, j} \gets \lnot X_{i,j}$
		\EndIf
	\EndFor
\EndFor
\State \Return $\mathbf{\tilde X}$
\EndProcedure
\end{algorithmic}
\end{algorithm}

A convenient property of differential privacy is \textit{post-processing} \cite{dwork2006differential}. If $\mathcal M: \mathcal X \to \mathcal F$ is an $\left( \varepsilon, \delta \right)$-differentially private randomized algorithm and $f: \mathcal F \to \mathcal F^\prime$ is an arbitrary mapping, then $f \circ \mathcal M: \mathcal X \to \mathcal F^\prime$ is $\left( \varepsilon, \delta \right)$-differentially private. We set $\mathcal M = \mathcal M_{rr}$ and define $f$ such that $f \left( \mathbf{\tilde X} \right) = \textproc{Impute} \left( \mathbf z, \mathbf{\tilde X}, \boldsymbol \mu, \boldsymbol \rho \right)$ for some fixed values $\mathbf z$, $\boldsymbol \mu$, and $\boldsymbol \rho$ (see below on the meaning of these parameters). Then by post-processing, it follows that each call to \textproc{Impute} on the perturbed reference panel $\mathbf{\tilde X}$ will satisfy $\varepsilon$-differential privacy. In other words, once $\mathbf{\tilde X}$ has been collected on the imputation server and perturbed so as to satisfy local differential privacy, an unlimited number of queries are able to be made by an algorithmic adversary without divulging any one haplotype's value at any one site with a high degree of certainty.

\subsection{HMM-based genotype imputation}

We will also use the following notation:
\begin{itemize}
	\item 0, 1, and $\varnothing$: the minor allele, major allele, and  constant denoting an unobserved site to be imputed;
	\item $n$ and $m$: the number of reference samples and reference markers;
	\item $\left[ n \right]$: the set of reference haplotypes, represented as the index set;
	\item $\mathbf X = \left[ \mathbf x_1, \mathbf x_2, \ldots, \mathbf x_m \right]^\intercal \in \left\{ 0, 1 \right\}^{m \times n}$: the reference panel haplotype sequences, equivalent to a real (and more, specifically, binary) matrix;
	\item $\left( z_k \right)_{k=1}^m \in \left\{ 0, 1, \varnothing \right\}^m$: the sequence corresponding to the observed target haplotype that, because it may include the missing site letter, is \textit{not}, strictly speaking, a real vector;
	\item $\left( \hat z_k \right)_{k=1}^m \equiv \mathbf{\hat z} \in \left[ 0, 1 \right]^m$: the sequence of imputed haplotype dosages, equivalent to a real vector;
	\item $\mathbf y \in \left[ n \right]^m$: the site-wise identities of the reference haplotypes from which $\mathbf z$ is descended;
	\item $\boldsymbol \rho \in \left[0, 1 \right]^{m}$: the recombination rates \cite{li2003modeling, li2009genotype} such that $\rho_i = P(y_{i+1} = j_2 | y_{i} = j_1)\quad j_2 \neq j_1$, meaning that $\boldsymbol \rho$ is equivalent to a real vector (we simply let $\rho_m = 0$ as a dummy value);
	\item $\boldsymbol \mu \in \left[0, 1 \right]^m$: the mutation rates \cite{li2003modeling, li2009genotype} such $\mu_i = P(z_i \neq X_{i, j} | y_i = j)$, meaning that $\boldsymbol \mu$ is equivalent a real vector;
	\item $\mathbf M = \left[ \mathbf m_1, \mathbf m_2, \ldots, \mathbf m_m \right]^\intercal \in \left[0, 1 \right]^{m \times n}$: the emission probabilities in matrix form such that $\mu_i = P(z_i \neq X_{i, j} | y_i = j)$ such that
		\begin{equation}
			M_{i,j} = \begin{cases}
							1 - \mu_i \quad &\text{if} \quad z_i = X_{i,j} \\
							\mu_i \quad &\text{if} \quad z_i = \varnothing \quad \text{or} \quad z_i \neq X_{i,j} 
						\end{cases};
		\end{equation}
	\item $\boldsymbol \Gamma = \left[ \boldsymbol \gamma_1, \boldsymbol \gamma_2, \ldots, \boldsymbol \gamma_m \right]^\intercal \in \left\{ 0, 1 \right\}^{m \times n}$: the posterior probabilities for haplotype identity for all sites in matrix form $\Gamma_{i,j} = P \left(y_i = j | \left( z_k \right)_{k=1}^m \right)$;
	\item $\mathbf A = \left[ \boldsymbol \alpha_1, \boldsymbol \alpha_2, \ldots, \boldsymbol \alpha_m \right]^\intercal \in \left\{ 0, 1 \right\}^{m \times n}$: the forward probabilities \cite{baum1972inequality} for all sites in matrix form such that $A_{i,j} = P(y_i = j | \left( z_k \right)_{k=1}^i)$;
	\item $\mathbf B = \left[ \boldsymbol \beta_1, \boldsymbol \beta_2, \ldots, \boldsymbol \beta_m \right]^\intercal \in \left\{ 0, 1 \right\}^{m \times n}$: the backward probabilities \cite{baum1972inequality} for all sites in matrix form such that $B_{i,j} = P \left(y_i = j, \left( z_k \right)_{k=i+1}^m \right)$;
\end{itemize}

By the Law of the Unconscious Statistician, we compute the expected value (i.e., dosage) for each $i$th site via the inner product of the posterior probabilities $\boldsymbol \gamma_i$ over the reference haplotype space $y_i$ and the values $\boldsymbol x_i$ that the reference haplotypes have at the $i$th site:
\begin{equation}
	\text{E} \left[ \hat z_i | \left( z_k \right)_{k=1}^m; \mathbf X, \boldsymbol \mu, \boldsymbol \rho  \right] = \mathbf x_i^\intercal \boldsymbol \gamma_i.
\end{equation}
As is critical to the forward-backward algorithm \cite{baum1972inequality}, we note that
\begin{equation}
	\boldsymbol \gamma_i \propto \boldsymbol \alpha_i \circ \boldsymbol \beta_i,
\end{equation}
where $\circ$ denotes the Hadamard product. However, $\boldsymbol \gamma_i$ must sum to unity, meaning that to compute the exact posterior distribution we normalize the two messages via
\begin{equation}
	\boldsymbol \gamma_i = \boldsymbol \alpha_i \circ \boldsymbol \beta_i \left( \boldsymbol \alpha_i^\intercal \boldsymbol \beta_i \right)^{-1}.
\end{equation}
For all sites in matrix form, this is equivalent to
\begin{equation}
	\text{E} \left[ \mathbf{\hat z} | \left( z_k \right)_{k=1}^m; \mathbf X, \boldsymbol \mu, \boldsymbol \rho  \right] = \left( \mathbf X \circ \boldsymbol \Gamma \right) \mathbf 1
\end{equation}
where
\begin{equation}
	\boldsymbol \Gamma = \left( \mathbf A \circ \mathbf B \right) \left( \text{diag} \left( \left( \mathbf A \circ \mathbf B \right)^\intercal \boldsymbol 1 \right) \right)^{-1}.
\end{equation}

As to computing the forward and backward messages, their recurrence relation from the Li-Stephens model \cite{li2003modeling, li2009genotype} can be shown to simplify, respectively, to
\begin{equation}
\begin{cases}
	\boldsymbol \alpha_1 = \mathbf 1 \\
	\boldsymbol \alpha_{i+1} = \mathbf{m}_{i+1} \circ \left( \frac{\rho_{i+1}}{n} \sum_{j=1}^n A_{i,j} + \left( 1 - \rho_i \right) \boldsymbol \alpha_{i} \right)
\end{cases}
\end{equation}
and
\begin{equation}
\begin{cases}
	\boldsymbol \beta_m = \mathbf 1 \\
	\boldsymbol \beta_i = \mathbf{m}_i \circ \left( \frac{\rho_i}{n} \sum_{j=1}^n B_{i+1,j} + \left( 1 - \rho_{i+1} \right) \boldsymbol{\beta}_{i+1} \right)
\end{cases}.
\end{equation}
This gives rise to the final implementation of \textproc{Impute}, which computes the forward and backward messages successively via dynamic programming \cite{bellman1966dynamic}.
\begin{algorithm}[h!]
\caption{Uses forward-algorithm to impute dosages according to Li-Stephens model.}
\begin{algorithmic}[1]
\Procedure{Impute}{$\mathbf z, \mathbf X, \boldsymbol \mu, \boldsymbol \rho$}
\State $\mathbf A, \mathbf B, \mathbf M \gets$ empty matrices
\For{$i = 1, 2, \ldots, m $}
	\For{$j = 1, 2, \ldots, n$}
		\If{$z_i = X_{i,j}$}
			\State $M_{i,j} \gets 1 - \mu_i$
		\Else
			\State $M_{i,j} \gets \mu_i$
		\EndIf
	\EndFor
\EndFor
\State $\boldsymbol \alpha_{1} \gets \boldsymbol 1$
\For{$i = 1, 2, \ldots, m $}
	\State $\boldsymbol \alpha_{i+1} = \mathbf{m}_{i+1} \circ \left( \frac{\rho_{i+1}}{n} \sum_{j=1}^n A_{i,j} + \left( 1 - \rho_i \right) \boldsymbol \alpha_{i} \right)$
\EndFor
\State $\boldsymbol \beta_{m} \gets \boldsymbol 1$
\For{$i = m - 1, m - 2, \ldots, 1 $}
	\State $\beta_i = \mathbf{m}_i \circ \left( \frac{\rho_i}{n} \sum_{j=1}^n B_{i+1,j} + \left( 1 - \rho_{i+1} \right) \boldsymbol{\beta}_{i+1} \right)$
\EndFor
\State $\boldsymbol \Gamma \gets \mathbf A \circ \mathbf B$
\State $\boldsymbol \Gamma \gets \boldsymbol \Gamma \left( \text{diag} \left( \boldsymbol \Gamma^\intercal \boldsymbol 1 \right) \right)^{-1}$
\State $\mathbf{\hat z} \gets \left( \mathbf X \circ \boldsymbol \Gamma \right) \mathbf 1$
\State \Return $\mathbf{\hat z}$
\EndProcedure
\end{algorithmic}
\end{algorithm}

\section*{Acknowledgements}
We thank Hoon Cho at the Broad Institute of MIT and Harvard for insightful discussion.

\bibliographystyle{IEEEtran}
\bibliography{shield}

\end{document}